\documentclass[aps,prl,preprint, showpacs]{revtex4}

\usepackage{epsfig}
\usepackage{graphicx}

\def\be{\begin{equation}}
\def\ee{\end{equation}}
\def\bea{\begin{eqnarray}}
\def\eea{\end{eqnarray}}

\begin{document}
\newcount\timehh  \newcount\timemm
\timehh=\time \divide\timehh by 60
\timemm=\time
\count255=\timehh\multiply\count255 by -60 \advance\timemm by \count255

\title{Orbital mechanisms of electron spin manipulation by an electric field}
\author{E. I. Rashba$^{1}$\cite{Rashba*} and Al. L. Efros$^2$}
\affiliation{$^1$Department of Physics, SUNY at Buffalo, Buffalo, New York 14260, USA\\
$^2$Naval Research Laboratory, Washington, DC 20375, USA}
\date{\today. SpinManipRevSubm.tex}

\begin{abstract}
A theory of spin manipulation of quasi-two-dimensional (2D) electrons by a time-dependent gate voltage applied to a quantum well is developed. The Dresselhaus and Rashba spin-orbit coupling mechanisms are shown to be rather efficient for this purpose. The spin response to a perpendicular-to-plane electric field is due to a deviation from the strict 2D limit and is controlled by  the ratios  of the spin, cyclotron and confinement frequencies.  The dependence of this response on the magnetic field direction is indicative of the strenghts of the competing spin-orbit coupling mechanisms. 
\end{abstract}

\maketitle


Manipulating electron spins is one of the central problems of the growing field of semiconductor spintronics\cite{reviews} that is of critical  importance for quantum computing and information processing\cite{LDV}. Most of the schemes proposed for computing with electron spins in quantum dots (QDs) and quantum wells (QWs) are based on using time-dependent magnetic fields. However for applications using time-dependent electric fields instead of magnetic ones would be highly preferable, and  various mechanisms of spin-orbit interaction \cite{RS91} open attractive possibilities for electrical control of electron spins. Recently Kato {\it et al.}\cite{Kato} successfully manipulated 2D electron spins by a gigahertz electric field. They used a parabolic Al$_{\rm x}$Ga$_{1-{\rm x}}$As QW formed by varying Al-content ${\rm x}={\rm x}(z)$ gradually across the well. The structure was specially engineered to achieve gate-voltage control of the $g$-factor through its dependence on ${\rm x}$ \cite{Salis}. Similar data on the electrical control of the $g$-factor were reported for GaAs/AlGaAs \cite{Yablonovitch}  and Si/SiGe \cite{J02} heterostructures. These achievements pave the way for manipulating electron spins in QDs individually.

Experimental success in achieving dynamical electric manipulation of electron spins raises the question about the dominant physical mechanisms controlling the coupling of spins to the electric field. The $\hat g$-tensor modulation resonance technique \cite{Kato} is based on the different dependence of the various $\hat g$-tensor components on the gate voltage and works when the external magnetic field ${\mbox{\boldmath $B$}}$ is tilted to the QW plane. Indeed, under these conditions the operators 
$(\mbox{\boldmath$\sigma$}{\hat g}{\mbox{\boldmath $B$}})$ and $(\mbox{\boldmath$\sigma$}{\hat g}^{\prime}{\mbox{\boldmath $B$}})$, where \mbox{\boldmath$\sigma$} is the Pauli matrix vector and ${\hat g}^{\prime}=d{\hat g}/dV$ is the derivative of ${\hat g}$ with respect to the gate voltage, do not commute. As a result, a time-dependent gate voltage $V(t)=V_0+{\tilde V}\sin(\omega t+\phi)$  leads to spin flip transitions at the spin resonance frequency $\omega=\omega_s=g\mu_BB/\hbar$, $\mu_B$ being the Bohr magneton.  The concept of the $\hat g$-tensor mechanism of spin-flip transitions suggested using the region of a very small $\hat g$-tensor, $\vert g\vert\alt 0.1$, where $\hat g$ is strongly anisotropic and gate-voltage dependent\cite{Kato}. 

In this letter we develop the theory for a different mechanism of gate-voltage induced spin resonance based on the electron's orbital motion and the standard mechanisms of spin-orbit coupling. This theory also requires a tilted magnetic field but does not require the $\hat g$-tensor to be small. Just the opposite, the large $g$ factors typical of narrow-gap A$_{3}$B$_5$ compounds are advantageous. Electric dipole spin resonance (EDSR) \cite{RS91} is especially strong when it is excited by an electric field lying in the QW plane. However, we show that it is also strong enough in the geometry when the time-dependent potential is applied to the gate, i.e., the time-dependent electric field is perpendicular to the well. This geometry is the most suitable for practical devices. Our results demonstrate convincingly that {\it efficient electrical spin manipulation can be achieved through the orbital mechanisms of spin-coupling to the electric field}. 

Two basic mechanisms of the spin-orbit coupling of 2D electrons are directly related to the symmetry properties of QWs. They stem from the structure inversion asymmetry (SIA) mechanism described by the Rashba term \cite{BR} and  the bulk inversion asymmetry (BIA) mechanism described by the Dresselhaus term \cite{D55,DT}. In GaAs QWs both terms are usually of the same order of magnitude \cite{GaAs} while in narrow-gap compounds like InAs the SIA mechanism dominates. Developing a reliable experimental technique based on EDSR requires a tool that identifies the spin-orbit mechanisms contributing to spin-flip transitions and allows to establish, as applied to specific materials, the dominating mechanisms. To this end, we find the EDSR intensity for the Dresselhaus and Rashba models as a function of the magnetic field direction. Our results suggest that {\it the angular dependence of the EDSR intensity is an unique characteristic of the various competing mechanisms of spin-orbit coupling  contributing to EDSR.}
 
In what follows we consider electrons confined in a parabolic QW  along the $z$ direction.  Then the electron Hamiltonian is $\hat{H}=\hat{H}_0+\hat{H}_Z+\hat{H}_{\rm so}$, where
\begin{equation}
\hat{H}_0={\hbar^2\hat{\mbox{\boldmath $k$}}^2\over 2m}+{m\omega_0^2z^2\over 2}~{\rm and}~\hat{H}_Z={1\over 2}\mu_B(\mbox{\boldmath $\sigma$}{\hat g}\mbox{\boldmath $B$})
\label{eq1}
\end{equation}  
are the orbital and Zeeman Hamiltonians, respectively.  Here $m$ is the electron effective mass, $\omega_0$ is the characteristic frequency of the parabolic potential, $\hat{\mbox{\boldmath $k$}}=-i\mbox{\boldmath $\nabla$}+e\mbox{\boldmath $A$}/\hbar c$, $\mbox{\boldmath $A$}$ is the vector-potential of the field $\mbox{\boldmath $B$}(\theta,\varphi)$, and $\theta$ and $\varphi$ are the polar and the azimuthal angles of $\mbox{\boldmath $B$}$. We have chosen parabolic confinement because it is known to be the only kind that can be solved exactly \cite{parab} for arbitrary $\mbox{\boldmath $B$}$ direction. The solution reveals the basic regularities of EDSR, including its dependence on the confinement strength. The spin-orbit interaction $\hat{H}_{\rm so}$ will be considered as a perturbation.

Because $\hat{H}_0$ has quadratic form both in the momenta and coordinates it can be diagonalized. Let us choose a new Cartesian frame where the $z'$-axis is parallel to $\mbox{\boldmath $B$}$ and the $y'$ axis is in the QW plane, the Landau gauge $\mbox{\boldmath $A$}=(0,Bx',0)$, and introduce new variables: $\xi=x'\cos\gamma-z'\sin\gamma$, and $\eta=x'\sin\gamma+z'\cos\gamma$. Then $\hat{H}_0$ can be written as sum of two harmonic oscillators:
\be
\hat{H}_0=\sum_{\zeta=\xi, \eta}(\hat{a}_\zeta \hat{a}^+_\zeta+\hat{a}_\zeta^+\hat{a}_\zeta)\hbar\omega_\zeta,~
\label{eq3}
\ee
where $\omega_\xi^2(\theta)=\omega_c^2\cos^2\gamma+\omega_0^2\sin^2(\theta+\gamma)$ and $\omega_\eta^2(\theta)=\omega_c^2\sin^2\gamma+\omega_0^2\cos^2(\theta+\gamma)$ are the frequencies, $E(n_\xi, n_\eta)=\sum_\zeta\hbar\omega_\zeta(n_\zeta+1/2)$ are the energy levels, and $n_{\xi,\eta}\geq 0$  \cite{parab}. Here $\omega_c=eB/mc$ is the cyclotron frequency for $\mbox{\boldmath $B$}\parallel{\bf {\hat{z}}}$, and $\gamma$ is determined by the decoupling condition $\sin 2\gamma=(\omega_0/\omega_c)^2\sin[2(\theta+\gamma)]$. The operators $\hat{a}_\zeta$ and $\hat{a}_\zeta^+$ are defined by the following relations: $\zeta-\zeta_0=\sqrt{\hbar/{2m\omega_\zeta}}(\hat{a}_\zeta^++\hat{a}_\zeta)$ and $\hat{k}_\zeta=i\sqrt{{m\omega_\zeta}/{2\hbar}}(\hat{a}_\zeta^+-\hat{a}_\zeta)$. The shifts in coordinates, $\xi_0$ and $\eta_0$, are related to the Landau momentum, $k\equiv k_{y'}$, as
$\xi_0\cos\gamma+\eta_0\sin\gamma=\lambda^2k$, where $\lambda=(\hbar c/eB)^{1/2}$ is the magnetic length. It is important that the operators of the kinetic momenta in the original frame, $({\hat k}_x, {\hat k}_y, {\hat k}_z)$, which are used in the following calculations, can be expressed as linear combinations of the operators $\hat{a}_\zeta$ and $\hat{a}_\zeta^+$. The coefficients depend only on the angles, $\theta$ and $\varphi$, and the frequencies, $\omega_\xi$ and $\omega_\eta$, and are independent of $\xi_0$, $\eta_0$, and $k$. The frequencies of the coupled cyclotron-confinement modes, $\omega_\zeta(\theta)$,  depend on the $\mbox{\boldmath $B$}$ direction. When $\omega_c<\omega_0$, $\omega_\xi(\theta)$ decreases from $\omega_c$ at $\theta=0$ to zero at $\theta=\pi/2$, while $\omega_\eta(\theta)$ increases from $\omega_0$ to $(\omega_0^2+\omega_c^2)^{1/2}$.

In presence of a high-frequency electric field, the term ${\hat H}_{\rm so}$ adds spin-orbit contributions to the time-dependent and time-independent parts of the total Hamiltonian $\hat{H}$. The latter contribution leads to mixing of the spin levels and it is convenient to eliminate it, in the first order in ${\hat H}_{\rm so}$, by a canonical transformation $\exp({\hat T})$\cite{RS91}. The operator $\hat T$ is non-diagonal in the orbital quantum numbers $(n_\xi, n_\eta)$, and its matrix elements are 
\be
\langle n_\xi^\prime, n_\eta^\prime, \sigma^\prime\vert {\hat T}\vert n_\xi, n_\eta, \sigma\rangle={{\langle n_\xi^\prime, n_\eta^\prime, \sigma^\prime\vert {\hat H}_{\rm so}\vert n_\xi, n_\eta, \sigma\rangle}\over{E_{\sigma^\prime}(n_\xi^\prime, n_\eta^\prime)-E_\sigma(n_\xi, n_\eta)}},
\label{eq8}
\ee
where $\sigma$ is the spin index. After the canonical transformation, the time-independent part of $\hat{H}$ conserves the electron spin projection on the magnetic field direction.

Because the motion in the direction of the time-dependent electric field, $\mbox{\boldmath $\tilde{E}$}(t)\parallel{\bf\hat{z}}$, is confined by the parabolic potential, the time-dependent interaction $ez\tilde{E}(t)$ is bounded. It does not depend on the spin, and the $z$-coordinate can be expressed in terms of $\hat{a}_\xi$ and $\hat{a}_\eta$
\bea
z&=&-\sqrt{\hbar/2m\omega_\xi}\sin(\theta+\gamma)(\hat{a}_\xi+\hat{a}^+_\xi)\nonumber\\
&&+\sqrt{\hbar/2m\omega_\eta}\cos(\theta+\gamma)(\hat{a}_\eta+\hat{a}^+_\eta).
\label{z}
\eea
The $\hat{T}$ transformation produces a commutator $\hat{z}_{\rm so}=[\hat{T},z]$; hence, $z$ acquires a spin-dependent part $\hat{z}_{\rm so}$. Spin-flip transitions are induced only by the spin-orbit contribution $e\hat{z}_{\rm so}{\tilde E}(t)$ to the time-dependent part of $\hat{H}$. 

To find the intensities of the spin-flip transitions excited by an electric field applied in the $z$ direction, one should calculate matrix elements of $\hat{z}_{\rm so}$ that are non-diagonal in spin projections and diagonal in the orbital quantum numbers. We will consider first the spin flip matrix element of $\hat{z}_{\rm so}$ in the case when spin-orbit interaction  $\hat{H}_{\rm so}$ is dominated by the Rashba term 
\be
\hat{H}_R=\alpha_R(\sigma_x\hat{k}_y-\sigma_y\hat{k}_x).
\label{eq9}
\ee
Of course, the parabolic confinement that is symmetric in $z$ does not produce the Rashba term by itself. Therefore, we introduce it phenomenologically, e.g., as originating from the hexagonal symmetry of a wurtzite type crystal. A reliable estimate of the spin-orbit coupling constant can be obtained today only  for the BIA mechanism that will be considered below. To simplify equations and elucidate the basic physics, we consider from now on the quantum limit, when only the lowest electron level $n_\xi=n_\eta=0$ is populated. 

A cumbersome algebra using several identities relating the frequencies  $\omega^2_{\xi,\eta}$ and the angles $\theta$ and $\gamma$ like $\omega_\xi^2\tan\gamma=\omega_\eta^2\tan(\theta+\gamma)$ results in a simple final equation for the matrix element of the spin-flip transition
\be
\langle\uparrow\vert \hat{z}_{\rm so}\vert\downarrow\rangle_R 
=-{\alpha_R\over{2\hbar}}
{{\omega_c\omega_s(\omega_c-\omega_s)\sin 2\theta}\over{(\omega_\xi^2(\theta)-\omega_s^2)(\omega_\eta^2(\theta)-\omega_s^2)}}~. 
\label{eq10}
\ee
The denominator of Eq.~(\ref{eq10}) can be rewritten explicitly as $[\omega_c^2\omega_0^2\cos^2\theta-\omega_s^2(\omega_0^2+\omega_c^2-\omega_s^2)]$. The factor $\sin2\theta$ that vanishes both for $\theta=0$ and $\theta=\pi/2$ reflects importance of a tilted magnetic field. 

The angular dependence of the EDSR intensity $I(\theta,\phi)\propto |\langle\uparrow\vert \hat{z}_{\rm so}\vert\downarrow\rangle_R|^2$ caused by the Rashba term is shown in Fig. 1$a$. Because of the poles of the denominator, the EDSR intensity increases when $\omega_s$ approaches one of the eigenfrequencies; practically, for $\omega_c<\omega_0$ only the pole $\omega_\xi(\theta)=\omega_s$ is important. In the strong confinement regime, when $\omega_c,~\omega_s\ll\omega_0$, $\omega_\xi\approx\omega_c\cos\theta$ and becomes the cyclotron frequency of 2D electrons in a tilted magnetic field. The sharpness of the resonance peak is cut-off  by a level width and also by the level anticrossing caused by the spin-orbit interaction.

When the spin-orbit Hamiltonian  $\hat{H}_{\rm so}$ is dominated by the bulk Dresselhaus spin-orbit interaction, the calculation of $\hat{z}_{\rm so}$ allows evaluating magnitudes of  the EDSR for specific A$_3$B$_5$ compounds. In the principal crystal axes, the 3D Dresselhaus spin-orbit Hamiltonian $\hat{\cal H}_D$ reads
\be
\hat{\cal H}_D=\delta(\mbox{\boldmath $\sigma$}\cdot\hat{\mbox{\boldmath $\kappa$}}),~~{\rm where}~~
{\hat\kappa}_x={\hat k}_y{\hat k}_x{\hat k}_y-{\hat k}_z{\hat k}_x{\hat k}_z~,
\label{eq11}
\ee 
 ${\hat\kappa}_y$ and ${\hat\kappa}_z$ can be derived by cyclic permutations, and $\delta$ is a parameter. We have found a general expression for the matrix element $\langle\uparrow\vert \hat{z}_{\rm so}\vert\downarrow\rangle_D$ for a [0,0,1] QW in a A$_{3}$B$_5$ crystal. The angular dependence of the EDSR intensity caused by the Dresselhaus term, $I(\theta,\phi)\propto |\langle\uparrow\vert \hat{z}_{\rm so}\vert\downarrow\rangle_D|^2$, is shown in Fig.~1$b$. The expression for $\langle\uparrow\vert \hat{z}_{\rm so}\vert\downarrow\rangle_D$ simplifies in the strong confinement limit:
\bea
\FL
&&\langle\uparrow\vert \hat{z}_{\rm so}\vert\downarrow\rangle_D 
\simeq{\delta m\over 2\hbar^2}
{\omega_c\omega_s\sin\theta\over \omega_0(\omega_\xi^2(\theta)-\omega_s^2)}\nonumber\\
&&\times\left[(\omega_c-\omega_s)\cos\theta\sin2\varphi-i(\omega_c\cos^2\theta-\omega_s)\cos2\varphi]\right].~~ 
\label{eq12}
\eea
This equation describes a quasi-2D regime when $\hat{\cal H}_D$ reduces to a 2D Dresselhaus term ${\hat H}_D=\alpha_D(\sigma_xk_x-\sigma_yk_y)$ with $\alpha_D=-\delta\langle k_z^2\rangle=-\delta m \omega_0/2\hbar$.

Eq.~(\ref{eq12}) has a pole at $\omega_c\cos\theta\approx\omega_s$, similarly to Eq.~(\ref{eq10}) in the strong confinement regime. A distinctive feature of the BIA mechanism is a strong azimuthal dependence of the EDSR intensity, $I(\theta, \varphi)$, that possesses a four-fold axis symmetry. Remarkably, the contribution of the Dresselhaus term does not vanishes for an in-plane magnetic field, $\theta=\pi/2$. In this geometry, the EDSR intensity does not depend on $\omega_s$, which drops from Eq.~(\ref{eq12}) because $\omega_\xi(\pi/2)=0$ and shows an especially strong azimuthal dependence on the magnetic field direction, $I(\pi/2, \varphi)\propto\cos^2 2\varphi$.

Figure 1 shows a drastic difference in the EDSR angular dependences caused by the BIA and SIA mechanisms. The four-fold symmetry of the EDSR angular dependence should be broken when the contributions from both mechanisms to the EDSR amplitude are of comparable magnitude (This is also true for breaking the symmetry of the energy spectrum \cite{RS91,GaAs}). Fig.~1 shows convincingly that the angular dependence of EDSR is a powerful tool for identifying contributions of the different competing mechanisms of spin-orbit coupling.

The efficiency of the BIA mechanism of EDSR can be evaluated by using the characteristic length, $l_D$, that is equal to the matrix element of $\hat{z}_{\rm so}$. Equation~(\ref{eq12}) gives $l_D\sim\delta m/2\hbar^2$ when all frequencies are of the same order of magnitude, $\omega_0\sim\omega_c\sim\omega_s$. We estimate $l_D\sim 10^{-9}$ to $10^{-8}$\,cm using a typical value $m\sim0.05m_0$ for the mass and also $\delta\approx$ 20\,eV\AA$^3$ for GaAs and 200\,eV\AA$^3$  for  InSb or GaSb \cite{delta}. It is much larger than the electron Compton length, $\lambdabar_C=\hbar/m_0c\approx 4\times 10^{-11}$\,cm, that plays the role of a characteristic length for EPR \cite{inter}. Therefore, {\it electrical manipulation of electron spins is preferable to magnetic not only because it allows access to the spins at a nanometer scale but also because a larger coupling constant can be achieved}.

However, there are several factors related to the electron confinement in a QW that can reduce $l_D$. It is seen from Eq.~(\ref{eq12}) that the confinement frequency $\omega_0$ appears in the denominator, therefore, $l_D$ includes a small factor $\omega_c/\omega_0\ll 1$. The factor $\omega_c/\omega_0$ reflects the deviation of the system from a strictly 2D geometry that is a critical condition for the gate-voltage controlled EDSR (the strict 2D limit corresponds to $\omega_0\rightarrow\infty$). This factor  was not really small in the Kato {\it et al.} experiment, $\omega_c/\omega_0\approx 0.5$,  because a wide parabolic well with effective width about 50\,nm and a strong magnetic field $B$ = 6T were used \cite{Kato}. In such a well $\alpha_D\approx 0.3\times 10^{-10}$\,eV\,cm which is much less than the typical value of the Rashba constant $\alpha_R\sim 10^{-9}$\,eV\,cm for InAs based QWs \cite{InAs}; and even larger values $\alpha_R\approx (3-6)\times 10^{-9}$\,eV\,cm were reported in Ref.\cite{alphaR}. This fact suggests that using asymmetric QWs should provide considerable advantages, and the corresponding length $l_R$ may be larger than $l_D$. However, specific calculations of $\alpha_R$ depend strongly on the boundary conditions \cite{alpha}, and the dependence of $\alpha_R$ on the QW width has not been investigated. 

Both $l_D$ and $l_R$ are also reduced because of the spin-flip frequency $\omega_s$ in the numerators of Eqs.~(\ref{eq10}) and (\ref{eq12}). It introduces a numerical factor $\omega_s/\omega_c = gm/2m_0$ that is about 0.16 in GaSb and InAs and about 0.32 in InSb. Therefore, usually $\omega_s/\omega_0$ rather than $\omega_c/\omega_0$ is the factor controlling the intensity of EDSR. It originates because of the parabolic confinement in the $z$ direction; a similar factor appears in the theory of EDSR for impurity centers \cite{RS91}. For an in-plane magnetic field, motion in the $\mbox{\boldmath $B$}$- direction becomes unrestricted, and that is why $\omega_s$ cancels in Eq.~(\ref{eq12}). In this case $l_D=(\delta m/2\hbar^2)(\omega_c/\omega_0)$. Therefore, the orbital mechanisms of spin-orbit coupling can provide a strong EDSR only if the ratio $\omega_s/\omega_c$ is not too small.

Another mechanism of EDSR, explored experimentally by McCombe {\it et al.}  \cite{MBK} for bulk InSb, is related to the anomalous coordinate    $\hat{\mbox{\boldmath $r$}}_{\rm so}=l^2_{\rm so}(\mbox{\boldmath $\sigma $}\times\hat{\mbox{\boldmath $k$}})$ introduced by Yafet \cite{YS}. The explicit expression for $l^2_{\rm so}$ in the framework of the Kane model is     $l_{\rm so}\approx\hbar(|g|/4m_0E_G)^{1/2}$, where $E_G$ is the forbidden gap \cite{YS,Sheka}. It is rather large, $l_{\rm so}\agt 10^{-8}$ cm. However, the operator $\hat{\mbox{\boldmath $r$}}_{\rm so}$ itself allows only the transitions at the combinational frequencies $\omega_\xi\pm\omega_s$, hence, it can produce spin-flip transitions only in the second order of perturbation theory, in combination with some different mechanism, e.g., QW asymmetry. As a result, the corresponding length $l_r$ is much smaller than  $l_{\rm so}$. For a strongly asymmetric QW, the upper bound for this length is $l_r\sim(\hbar\omega_s/E_G)L_{\rm conf}$, where $L_{\rm conf}=\sqrt{\hbar/m\omega_0}$ is the confinement length. Unfortunately, we are not aware of any more specific experimental or numerical data on $l_r$. 

The spatial dependence of the electron $\hat g$-tensor across the QW that allows efficient electric control of the Zeeman splitting \cite{Kato,Salis} also results in a spin-orbit coupling because $\hat{H}_Z=\mu_B(\mbox{\boldmath $\sigma $}\hat{g}(z)\mbox{\boldmath $B$})/2$ includes both the coordinate $z$ and Pauli matrices. The magnitude of the electron spin coupling to the time dependent voltage $V(t)$ is about $\sim \mu_B{\tilde V}(dg/dV) B$, where $B=6$\,T. The derivative $dg/dV$ is about $dg/dV\sim g/V_0$ with $g\sim 0.1$ and $V_0\sim 1$V, see Fig.\,2B of Ref.~\onlinecite{Kato}.  With ${\tilde V}\approx {\tilde E}w$  and $w\approx 100$\,nm, the characteristic length $l_g$ caused by the spatial dependence of $\hat{g}$ is $l_g\sim \lambdabar_CgwB/V_0\approx 7\times 10^{-10}$\,cm. It should be compared with $l_D$ that turned out to be negligibly  small, $l_D\alt 10^{-11}$\,cm, due to  anomalously small $g$-factor in GaAs (the ratio $\omega_s/\omega_c=gm/2m_0\alt 10^{-2}$). Therefore, our estimates show that the gate-voltage manipulation of the electron spins achieved by Kato {\em et al.} \cite{Kato} was performed with a characteristic spin-orbit length $l$ at the level of $l\approx 7\times10^{-10}$ cm.  

In most narrow gap semiconductor compounds with their typically large $g$-factors, and especially in those with strong spin-orbit coupling, the BIA and SIA orbital mechanisms dominate the coupling of electron spins to a perpendicular electric field in moderate and strong magnetic fields, and the stength of this coupling is sufficient for electron spin manipulation. Generally, however, relative contributions of the various spin-flip transition mechanisms depend strongly on the specific semiconductor material \cite{Si} and the geometry of the QW.

In conclusion, we have shown that the dynamic spin response to an electric field perpendicular to the QW plane is controlled by the deviation of the confined electrons from strictly 2D behavior. Therefore, the response of the spin system to the gate voltage depends strongly on the ratio of the confinement layer thickness to the magnetic length, that should not be too small. Semiconductor compounds with large $g$-factors are highly advantageous for gate-voltage driven EDSR. Mixing the in-plane and perpendicular orbital motion is  critical for EDSR, and for a (0,0,1) QW it requires that the magnetic field is tilted. The dependence of the spin-resonance intensity on the magnetic field direction with respect to the crystal axes is indicative of the role of the various mechanisms of spin-orbit coupling involved. The Dresselhaus and Rashba spin-orbit coupling  mechanisms are efficient enough for electron spin manipulation by a time-dependent gate voltage.

E.I.R. and Al.L.E.  acknowledge the financial support from DARPA/SPINS by the ONR Grant N000140010819 and from DARPA/QuIST and ONR, respectively, and thank T. Kennedy for closely reading the manuscript.

\begin{figure}[th]
\caption{Angular dependence of the EDSR intensity $I(\theta, \varphi)$ for a (0,0,1) QW (in arbitrary units) calculated for (a)-- Rashba SIA and (b)--Dresselhaus BIA mechanisms. Parameter values $\omega_s/\omega_c=-0.17$ (as in InAs) and $\omega_c/\omega_0=0.5$ (as in Ref.~\cite{Kato}) were used for both mechanisms.  To cut-off the pole in $I(\theta, \varphi)$ at $\omega^2_\xi(\theta)=\omega_s^2$, an imaginary part $i\Gamma$ with $\Gamma=0.08\omega_c$ was added to $\omega_c$.}
\label{angular}
\end{figure}

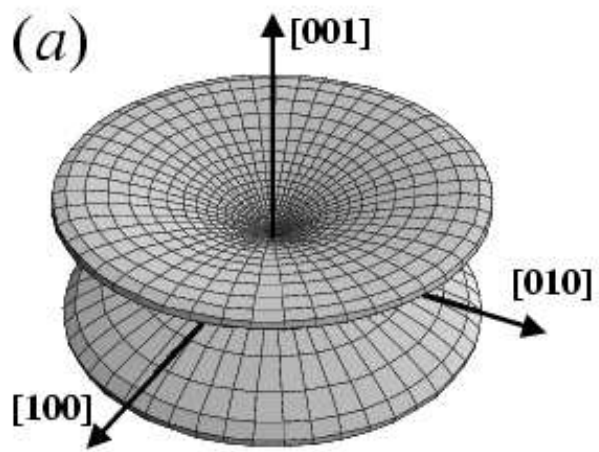
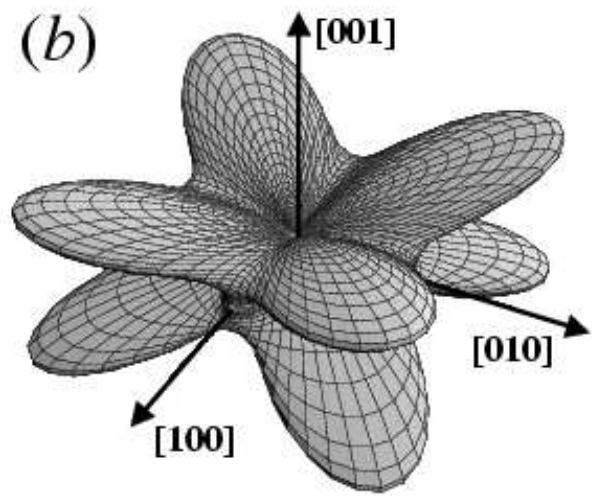
\end{document}